\documentclass[conference]{IEEEtran}
\IEEEoverridecommandlockouts
\usepackage{cite}
\usepackage{amsmath,amssymb,amsfonts}
\usepackage{algorithmic}
\usepackage{graphicx}
\usepackage{textcomp}
\usepackage{xcolor}
\usepackage[algo2e,linesnumbered,ruled]{algorithm2e}
\def\BibTeX{{\rm B\kern-.05em{\sc i\kern-.025em b}\kern-.08em
    T\kern-.1667em\lower.7ex\hbox{E}\kern-.125emX}}
\begin{document}

\title{Data-Driven 3D Placement of UAV Base Stations for Arbitrarily Distributed Crowds\\
\thanks{This research is supported by Ministry of Science and Technology under the Grant MOST 108-2634-F-009-006- through Pervasive Artificial Intelligence Research (PAIR) Labs, Taiwan.}
}

\author{\IEEEauthorblockN{Chuan-Chi Lai\IEEEauthorrefmark{1}, Li-Chun Wang\IEEEauthorrefmark{1}, and Zhu Han\IEEEauthorrefmark{2}}
	\IEEEauthorblockA{\IEEEauthorrefmark{1}Department of Electrical and Computer Engineering, National Chiao Tung University, Hsinchu 300, Taiwan\\
		\IEEEauthorrefmark{2}Department of Electrical and Computer Engineering, University of Houston, Houston, TX 77004, USA\\
		Email: cclai1109@nctu.edu.tw; lichun@cc.nctu.edu.tw; hanzhu22@gmail.com
	}%
}

\maketitle

\begin{abstract}
In this paper, we consider an Unmanned Aerial Vehicle (UAV)-assisted cellular system which consists of multiple UAV base stations (BSs) cooperating the terrestrial BSs. In such a heterogeneous network, for cellular operators, the problem is how to determine the appropriate number, locations, and altitudes of UAV-BSs to improve the system sumrate as well as satisfy the demands of arbitrarily flash crowds on data rates. We propose a data-driven 3D placement of UAV-BSs for providing an effective placement result with a feasible computational cost. The proposed algorithm searches for the appropriate number, location, coverage, and altitude of each UAV-BS in the serving area with the maximized system sumrate in polynomial time so as to guarantee the minimum data rate requirement of UE. The simulation results show that the proposed approach can improve system sumrate in comparison with the case without UAV-BSs.
\end{abstract}


\section{Introduction}
\label{intro}
In recent years, \emph{Unmanned Aerial Vehicle mounted Base Stations} (UAV-BSs) have become a new promising solution for providing temporary communication services to recover the disaster area or to satisfy the sudden demands (hot-spots) caused by \emph{Flash Crowds}, which is commonly referred to as \emph{UAV-Assisted Communications}~\cite{7744808}~\cite{8038869}. The advantage of using UAV-BSs is the flexible ability to provide dynamic and on-demand communications. The high altitudes of UAV-BSs enables them to effectively establish line-of-sight (LoS) communication links and mitigate signal blockage and shadowing. Accordingly, UAV-BSs become an agile solution to serve ground users arbitrarily distributed in a terrestrial infrastructure-less area. Compared to the deployment of traditional ground base stations (GBSs), deploying UAV-BSs is a cost-effective and energy-efficient solution which can save a significant amount of land cost for the cellular operators. 

Due to the characteristics of wireless propagation, there is a relation between the altitude and the optimal coverage of a UAV-BS, which is modeled in~\cite{6863654}. The authors modeled the air-to-ground (ATG) channel with derivations of the probabilities of LoS and non-line-of-sight (NLoS) signals and now their proposed channel model has been widely used in UAV communications. In consideration of the path loss constraint and uniform users in different environments, the optimal altitude, coverage, and location single deployed UAV-BS are discussed in~\cite{7510820}. In addition to the single UAV-BS case, many researchers have focused on the issues of 3D placement of multiple UAV-BSs. A spiral placement algorithm was proposed by~\cite{7762053} and it deployed multiple UAV-BSs with the fixed altitude and transmit power at optimal locations and minimize the number of deployed UAV-BSs to covered all users while considering various user densities. However, this approach only consider the fixed coverage and altitude of UAV-BS for the placement. 

Unfortunately, all the above conventional works only considered research issues from the perspectives of users and did not consider the coexistence of ground cellular systems. These related works also did not consider the arbitrary distribution of users for flash-crowd events, such as outdoor concerts, marathons, election campaigns. They only observed the system performance under some traditional stochastic user distribution models, such as uniform, Gaussian, and Poisson Point Process (PPP). Furthermore, most of them did not consider the coexistence of GBSs, the effects of co-channel interference, and the sumrate optimization problem. In practice, there are many challenging open issues for establishing such a UAV-assisted cellular system. In particular, the managing mechanisms of placement, resource allocation, power control, and flight scheduling are the urgent technologies to allow the deployed UAV-BSs to coexist with the terrestrial cellular systems. Such a UAV-assisted cellular system will be one important use case of 5G or beyond 5G networks, which is capable to serve dynamic traffic demands~\cite{7762185}. The effective and efficient technologies of UAV-BS placement/management thereby become popular topics in the communications domain. Hence, we focus on the dynamic placement of UAV-BSs over a terrestrial cellular system while improving the system performance in terms of the sumrate.

In this work, we discuss how to deploy and determine the appropriate altitude and location of each deployed UAV-BS to serve ground user equipments (UEs) in consideration of maximizing the system sumrate. Focusing on the downlink transmission from the GBS to UEs and from the UAV-BSs to their corresponding UEs, we propose a data-driven 3D placement algorithm which solves the considered placement problem efficiently. The proposed method firstly analyzes the distribution and density of ground UEs and then finds the possible candidate placements for clustered UEs. After that, the proposed method re-tunes the candidate altitude, location, and coverage of each UAV-BS in the considered area for maximizing the system sumrate with the optimization constraints on the co-channel interference and the allocated data rate of each UE.
The proposed adaptive algorithm can find an appropriate value of $k$ to deploy UAV-BSs with balanced serving loads (or served users). The simulation results indicate that the proposed approach can jointly satisfy the UE demands and provide a higher the system sumrate in comparison with the solution without UAV-BSs.


The balance of this paper is organized as follows. In Section~\ref{sec:problem} presents the considered system model, assumptions, and problem statement. Section~\ref{sec:3dplacement} introduces the proposed data-driven approach and a breakdown of the algorithms. 
Simulation results are presented in Section~\ref{sec:simulation}. Finally, we make concluding remarks in Section~\ref{sec:conclusion}.

\section{System Model and Problem Description}
\label{sec:problem}

\subsection{System Model}
\label{sec:systemmodel}
\begin{figure}[t]
	\centering
	\includegraphics[width=0.36\textwidth]{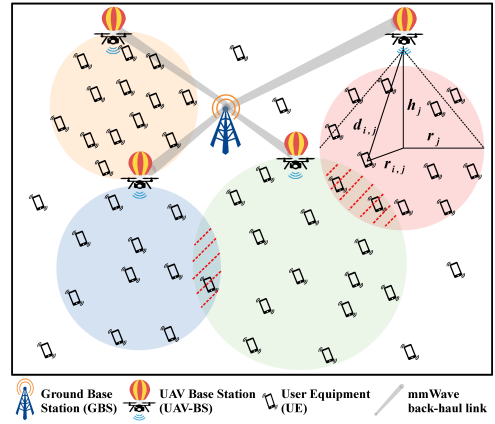}
	\caption{Architecture of the considered UAV-assisted cellular system.}
	\label{fig:system}
	\vspace{-15pt}
\end{figure}

As shown in Fig.~\ref{fig:system}, we consider a UAV-Assisted cellular system consisting one GBS, $G$, and a set of UAV-BSs, $\mathcal{U}=\{U_1, U_2, \dots, U_K\}$, in an urban scenario, where $K$ is the maximum number of available UAV-BSs. The UAV-Assisted cellular system serves a set of UEs, $E=\{u_1,u_2,\dots,u_N\}$, and the total number of UEs is $|E|=N$. The UAV-BSs can move in the sky to any position. In this paper, we consider the downlink transmissions. Each UE only uses the resource of one BS (GBS or UAV-BS) at a certain time. We assume that all the UEs are arbitrarily distributed on the ground due to the operation requirements, the terrain limitations, or unpredictable events. The placement decision of UAV-BSs is controlled by the edge controller (or controller) behind the GBS. All the UAV-BS and UEs are equipped with directional antennas to transmit and receive 4G LTE-A signals in the considered environments. We assume that the GBS and each UAV-BS use the same spectrum and provide the same bandwidth $B$ for the down links in the considered system. The GBS are also equipped a mmWave directional antenna array using another dedicated spectrum to provide an additional network volume for the back-haul communication link between the GBS and each UAV-BS. 

In our work, we focus on the downlink transmissions, and we introduce the radio propagation models for the downlink transmissions which consists of following three cases: 1) GBS to UAV-BS, 2) UAV-BS to UE, and 3) GBS to UE. We now respectively introduce these cases under the assumption that the appropriate number of deployed UAV-BSs is $k$, where $1\leq k\leq K$.

\subsubsection{GBS-to-UAV-BS Propagation Model}
In the considered system model, the GBS uses directional mmWave antennas to transmit signals to the UAV-BSs. The UAV-BSs fly at relatively high altitudes, so that the GBS-to-UAV-BS channel will be the simplest path-loss model using LoS links between the GBS and UAV-BSs which propagation in free space. Thus, the received signal power of each UAV-BS $U_j$ will be

{\small\vspace{-4pt}
\begin{equation}\label{eq:receive_power:uav}
P_j^{R}=P_G^{\textrm{mmWave}} g_G^T g_j^R \left(\dfrac{c}{4\pi d_{j,G}f_c^\textrm{mmWave}}\right)^2,
\vspace{-4pt}
\end{equation}}%
where $P_G^{\textrm{mmWave}}$ is the fixed transmit power of the mmWave antenna, $g_G^T$ is the transmit antenna gain of the GBS, $g_j^R$ is the receive antenna gain of $U_j$, $c$ is the speed of light, $f_c^\textrm{mmWave}$ is the used carrier frequency of the back-haul link, $d_{j,G}$ is the distance between the GBS and $U_j$, and $j=1,2,\dots,k$. According to~\eqref{eq:receive_power:uav} and the Shannon theorem, the back-haul capacity of a UAV-BS $U_j$ can be obtained by

{\small\vspace{-4pt}
\begin{equation}\label{eq:backhaul_rate:uav}
\hat{C}_j=B_{j,G}^\textrm{mmWave}\log_2\left(1+\dfrac{P_j^{R}}{B_{j,G}^\textrm{mmWave}N_0}\right),
\vspace{-4pt}
\end{equation}}%
where $B_{j,G}^\textrm{mmWave}$ is the allocated bandwidth (in Hz) of the mmWave back-haul link for UAV-BS $U_j$, and $N_0$ is the thermal noise power spectral density.

\subsubsection{UAV-BS-to-UE Propagation Model}
The second propagation model is used to model the downlink transmission from a UAV-BS to a UE.
Such a radio propagation model is well-known as the air-to-ground propagation channel and commonly modeled by considering the LoS and NLoS signals along with their occurrence probabilities separately~\cite{7417609}. We adopt the air-to-ground channel model in~\cite{6863654}, and the probabilities of LoS and NLoS for a UE $u_i$ associated with UAV-BS $U_j$ are

{\small\vspace{-10pt}
\begin{align}\label{PNLoS_to_user}
P_{h_j,r_{i,j}}^\text{LoS}&=\dfrac{1}{1+a\exp\left(-b(\dfrac{180}{\pi}\tan^{-1}(\dfrac{h_j}{r_{i,j}})-a)\right)},\nonumber\\
P_{h_j,r_{i,j}}^\text{NLoS}&=1-P_{h_j,r_{i,j}}^\text{LoS},
\end{align}}%
where $h_j$ is the altitude of each UAV-BS $U_j$, $a$ and $b$ are environment variables, $r_{i,j}$ is the horizontal euclidean distance between $u_i$ and $U_j$. Note that $r_{i,j}=\sqrt{(x_j-x_i)^2+(y_j-y_i)^2}$, $(x_j,y_j)$ is the horizontal location of UAV-BS $U_j$, $(x_{i},y_{i})$ is the horizontal location of UE $u_i$, $i=1,2,\dots,N$, and $j=1,2,\dots,k$. Considering the free space propagation loss, the channel model \cite{6863654} of the LoS and NLoS links can be written as

{\small\vspace{-15pt}
\begin{align}
L_{h_j,r_{i,j}}^\text{LoS}&=20\log\left(\dfrac{4\pi f_c d_{i,j}}{c}\right)+\eta_{LoS}, \nonumber\\
L_{h_j,r_{i,j}}^\text{NLoS}&=20\log\left(\dfrac{4\pi f_c d_{i,j}}{c}\right)+\eta_{NLoS},\label{model_nlos}
\end{align}}%
where $\eta_{LoS}$ and $\eta_{NLoS}$ are the mean additional losses for LoS and NLoS, $f_c$ is the carrier frequency of front-haul link, and $d_{i,j}=\sqrt{r_{i,j}^2+h_j^2}$ is the euclidean distance between $u_i$ and $U_j$. According to~\eqref{PNLoS_to_user} and~\eqref{model_nlos}, and let $\theta_{i,j}={\tan^{-1}}\frac{h_j}{r_{i,j}}$, we can obtain the average ATG channel model between $u_i$ and $U_j$ and it is denoted as

{\small\vspace{-10pt}
\begin{align}\label{average_atg_model}
L_{h_j,r_{i,j}}&=P_{h_j,r_{i,j}}^\text{LoS}L_{h_j,r_{i,j}}^\text{LoS}+P_{h_j,r_{i,j}}^\text{NLoS}L_{h_j,r_{i,j}}^\text{NLoS} \nonumber\\
&=\dfrac{\eta_{LoS}-\eta_{NLoS}}{1+a\exp\left(-b(\dfrac{180}{\pi}\tan^{-1}\theta_{i,j}-a)\right)} \nonumber\\
&+20\log({r_{i,j} \sec\theta_{i,j}})+20\log(\dfrac{4\pi f_c}{c})+\eta_{NLoS}.
\end{align}\vspace{-15pt}}%

Let $P_{i,j}$ be the minimum required transmit power for transmitting signal from UAV-BS $U_j$ to UE $u_i$, the transmission is successful if the received signal-to-interference-plus-noise ratio (SINR) at a UE is larger than a certain threshold $\gamma_{\textrm{th}}$. Thus, SINR expression for a UE $u_i$ associated with UAV-BS $U_j$ is 

{\small\vspace{-10pt}
\begin{equation}\label{eq:sinr:d2u}
\gamma_{i,j}=\frac{P_{i,j}(10^{L_{h_j,r_{i,j}}/10})^{-1}}{I_G+I_{\mathcal{U}\setminus\{U_j\}}+B_{i,j}N_0},
\end{equation}}%
where $I_G$ is the received interference power from the GBS and $I_{\mathcal{U}\setminus\{U_j\}}=\sum_{j'=1}^{k}P_{i,j'}(10^{L_{h_j,r_{i,j'}}/10})^{-1}\psi_{j,j'}$ is the interference power from the nearby UAV-BSs if UE $u_i$ locates in the overlapped coverage, where $\psi_{j,j'}=1$ if $u_i$ locates in the overlapping coverage area of UAV-BSs $U_j$ and $U_{j'}$, and $U_{j'}\in\mathcal{U}, \forall j'\neq j$; otherwise, $\psi_{j,j'}=0$.
According to the Shannon theorem and~\eqref{eq:sinr:d2u}, the allocated data rate (in bps) of $u_i$ associated with $U_j$ will be

{\small\vspace{-5pt}
\begin{equation}\label{eq:data_rate:uav2ue}
c_{i,j}=B_{i,j}\text{Pr}\{\gamma_{i,j}>\gamma_{\textrm{th}}\}\log_2(1+\gamma_{i,j}),
\end{equation}}%
where $B_{i,j}$ is the allocated bandwidth (in Hz) of down-link connection from UAV-BS $U_j$ to a served UE $u_i$. The transmit power allocated to $u_i$ of interest can be obtained by

{\small\vspace{-5pt}
\begin{equation}\label{eq:transmit_power:uav_to_1_ue}
P_{i,j}=10^{L_{h_j,r_{i,j}}/10}(I_G+I_{\mathcal{U}\setminus\{U_j\}}+B_{i,j}N_0)(2^{c_{i,j}/B_{i,j}}-1).
\end{equation}}%
Then, the potential total transmit power of UAV $U_j$ for serving its associated UEs can be calculated as
\begin{equation}\label{eq:transmit_power:uav_to_all_ue}
P_j=\sum_{i=1}^{N_j}P_{i,j},
\end{equation}
where $N_j$ is the number of UEs associated with the UAV-BS $U_j$. According to~\eqref{eq:data_rate:uav2ue}, the data transmission rate of the UAV-BS $U_j$ for serving its associated UEs is\\

\vspace{-15pt}
{\small
\begin{equation}\label{eq:sum_rate_constrait:uav_to_all_ue}
C_j=\sum_{i=1}^{N_j}c_{i,j}.
\end{equation}}%

\subsubsection{GBS-to-UE Propagation Model}
For terrestrial wireless channel between points $p_1$ and $p_2$, we consider a standard power law path-loss $L_{p_1,p_2}=||p_1-p_2||^{-\alpha}$ with path-loss exponent $\alpha>2$. All the terrestrial propagation signals are assumed to experience independent Rayleigh fading. The GBS are assumed to transmit at fixed power $P_G$ for terrestrial communications. The received power of UE $u_i$ served by the GBS is therefore $P_Ghr_{i,G}^{-\alpha}$, where $h\sim\exp(1)$ models Rayleigh fading and $r_{i,G}$ is the horizontal distance between a UE and the GBS. Since there are $k$ UAV-BSs in the considered system, the co-channel interference power experienced
by a UE can be expressed as

{\small\vspace{-5pt}
\begin{equation}\label{eq:interference:uav2ue}
I_{U}=\sum_{j=1}^{k}P_jhr_{i,j}^{-\alpha},
\end{equation}}%
where $P_j$ is the transmit power of UAV-BS $U_j$ and $r_{i,j}$ is the distance from UE $u_i$ to UAV-BS $U_j$.
The SINR expression for a user $u_i$ that can connect to the GBS is

{\small\vspace{-5pt}
\begin{equation}\label{eq:sinr:g2u}
\gamma_{i,G}=\frac{P_Ghr_{i,G}^{-\alpha}}{I_{U}+B_{i,G}N_0},
\end{equation}}%
where $I_\mathcal{U}=\sum_{\forall U_{j}\in\mathcal{U}}P_{j}(10^{L_{h_j,r_{i,j}}/10})^{-1}$ is the total interference power from the other disassociated UAV-BSs. The achievable data rate (in bps) of a UE associated with the GBS can be calculated as

{\small\vspace{-5pt}
\begin{equation}\label{eq:rate:gbs_ue}
c_{i,G}=B_{i,G}\text{Pr}\{\gamma_{i,G}>\gamma_{\textrm{th}}\}\log_2(1+\gamma_{i,G}),
\end{equation}}%
where $B_{i,G}$ is the allocated bandwidth (in Hz) to $u_i$ associate with the GBS. The potential transmission rate (in bps) of the GBS can be obtained by~\cite{6042301}

{\small\vspace{-5pt}
\begin{equation}\label{eq:rate:gbs}
C_G=\frac{\lambda_G}{\pi r_G^2}\overline{c_{i,G}}=\sum_{i=1}^{N_G}c_{i,G},
\end{equation}}%
where $r_G$ is the coverage radius of the GBS, $\lambda_G$ is the UE density of GBS's service coverage, $\overline{c_{i,G}}$ is the average data rate of a UE associated with the GBS, and $N_G$ is the number of UEs which is associated with the GBS.

\subsection{Problem Formulation}
\label{sec:problem:formulation}
We focus on the case of deploying $k$ UAV-BSs in the target area to improve the downlink sumrate of the UAV-assisted cellular system with one GBS. The system model is depicted in Fig.~\ref{fig:system}. The considered decision problem of 3d UAV placement can be defined as follows.

Suppose that the notations and assumptions are defined as above, the considered problem is to search for the appropriate placement parameters $(x_j,y_j,h_j,r_j)$ of each UAV-BS $U_j$ with the minimized number of UAV-BSs $k$, $0\leq k\leq K$, such that
\begin{align}\label{eq:system_sum_rate}
\max_{x_j,y_j,h_j,r_j}&\enspace\sum_{i=1}^{N}c_{i,G}\delta_{i,G}+\sum_{j=1}^{k}\sum_{i=1}^{N}c_{i,j}\delta_{i,j},&\tag{P1}\\
s.t.&\enspace r_j\leq r_{\max}(h_j),&\label{eq:system_sum_rate:c1}\\
&\enspace h_{\min}\leq\enspace h_j\leq h_{\max},&\label{eq:system_sum_rate:c2}
\end{align}
\begin{align}
&\enspace c_{i,j}\delta_{i,j}+c_{i,G}\delta_{i,G}\geq c_{\min}, i=1,2,\dots,N,\nonumber\\
&\hspace{11.25em} j=1,2,\dots,k,&\label{eq:system_sum_rate:c3}\\
&\enspace \sum_{i=1}^{N}c_{i,G}\delta_{i,G}\leq \hat{C}_G, &\label{eq:system_sum_rate:c4}\\
&\enspace \sum_{i=1}^{N}c_{i,j}\delta_{i,j}\leq \hat{C}_j, \qquad\qquad j=1,2,\dots,k,\label{eq:system_sum_rate:c5}\\
&\enspace \sum_{i=1}^{N}\delta_{i,G}+\sum_{j=1}^{k}\sum_{i=1}^{N}\delta_{i,j}=N.\label{eq:system_sum_rate:c6}
\end{align}
where two indicator functions $\delta_{i,j}$ and $\delta_{i,G}$ are defined as
\begin{equation}\label{eq:indicator_for_association}
\delta_{i,j}=\begin{cases}
1, & \textrm{if $\gamma_{i,j}>\gamma_{\textrm{th}}$;}\\
0, & \textrm{otherwise},
\vspace{-5pt}
\end{cases}
\end{equation}
and $\delta_{i,G}=1-\sum_{j=1}^{k}\delta_{i,j}.$

In the considered problem~\eqref{eq:system_sum_rate}, the maximum coverage of $U_j$, $r_{\max}(h_j)$, in constraint~\eqref{eq:system_sum_rate:c1}, is determined by $h_j$, and the relation between the altitude and maximum coverage of a UAV-BS has been discussed in~\cite{6863654}. In constraint~\eqref{eq:system_sum_rate:c2}, the deployed altitude $h_j$ of each UAV-BS is only allowed within $[h_{\min},h_{\max}]$ which depends on the limitations of local laws and ability of the UAV. We also consider the demands of the minimum data rate from the cellular operator's aspect, and we define an admin parameter $c_{\min}$ for each UAV-BS $U_j$ or the GBS $G$ to guarantee the minimum allocated date rate of a UE in constraints~\eqref{eq:system_sum_rate:c3}. Constraint~\eqref{eq:system_sum_rate:c4} guarantees that the total downlink transmission rate of the links from the GBS and its associated UEs do not exceed the maximum ability of providing data rate $\hat{C}_G$. Constraint~\eqref{eq:system_sum_rate:c5} is used to make the total downlink transmission rate of the links from UAV-BS $U_j$ to its associated UEs do not exceed the maximum allocated data rate of back-haul link on $U_j$ according to~\eqref{eq:backhaul_rate:uav}. Constraint~\eqref{eq:system_sum_rate:c6} makes each UE only be associated with one UAV-BS or the GBS at a time.
The indicator function $\delta_{i,j}$ in~\eqref{eq:indicator_for_association} is used to indicate UE $u_i$ is associated with UAV-BS $U_j$ if the $\gamma_{i,j}>\gamma_{\textrm{th}}$, where $\gamma_{\textrm{th}}$ is a given SINR threshold.

\section{Data-Driven 3D Placement of UAV-BSs}
\label{sec:3dplacement}
In this section, we proposed a data-driven placement for improving the sumrate performance of the UAV-assisted cellular system in a more reasonable way, especially for the unpredictable events or flash crowds with arbitrary distributed users. Algorithm~\ref{alg:uav_placement} shows the pseudo-code of the proposed placement procedures. In addition, we describe the notations/variables in an in-text manner and use some comment texts to help the ease of understanding. The detailed explanations will be presented in following subsections.

\begin{algorithm2e}[!ht]
	\scriptsize
	\SetAlgoLined
	\KwIn{dataset of UE locations $L_E$, location of the GBS $L_G$, the maximum number of UAV-BS $K$, SINR threshold $\gamma_{\textrm{th}}$, the channel bandwidth provided by each UAV and the GBS $B$, the transmit power of the GBS $P_G$, the transmit power of a UAV-BS $P_{\textrm{uav}}$, and the minimum data rate requirement $c_{\min}$}
	\KwOut{association information $L_{\textrm{association}}$, UAV locations $L_{\textrm{uav}}^{\textrm{cand}}$, and UAV altitudes $L_{\textrm{altitude}}$}
	create a list $L_{\textrm{uav}}^{\textrm{cand}}$ to store the UAV locations\;
	$N\leftarrow L_E.length$\;
	create a list $D_G$ to record the distance between the GBS and each UE\;
	create a list $S_G$ to save the received power from the GBS on each UE\;	
	create a list $L_{\textrm{association}}$ to save the association information of each UE\;	
	create a list $L_{\textrm{SINR}}$ to save SINR corresponding to its associated UAV\;	
	create a list $L_{\textrm{radius}}$ to save the candidate coverage radius of each UAV\;	
	create a list $L_{\textrm{altitude}}$ to save the candidate altitude of each UAV\;
	\For{$i=1$ to $N$}{
		$D_G[i]\leftarrow\sqrt{(L_E[i].x-L_G.x)^2+(L_E[i].y-L_G.y)^2}$\;
		\tcc{$h$ modeles Rayleigh fading and the path-loss exponent $\alpha>2$}	
		$S_G[i]\leftarrow P_Gh*(D_G[i])^{-\alpha}$\;
		compute SINR $\gamma_{i,G}$ by~\eqref{eq:sinr:g2u} with the interference $I_U=0$ (mW)\;
		\If{$\gamma_{i,G}>\gamma_{\textrm{th}}$}{
			\tcc{The association value is $0,1,\dots,k$}
			$L_{\textrm{association}}[i]\leftarrow 0$\;
		}
	}
	get $N_G$ by checking the number of '0' in $L_{\textrm{association}}$\; 
	initialize $k$ by~\eqref{eq:initial_value_of_k}\;
	run balanced $k$-means clustering~\cite{AAAI1816711} to cluster the unassociated UEs with $L_G$ and update $L_{\textrm{association}}$\label{alg:uav_placement:19}\;
	\Repeat{all the uav locations in $L_{\textrm{uav}}^{\textrm{cand}}$ do not change\label{alg:uav_placement:20}}{
		do the placement refinement by finding the minimum covering circle of each cluster~\cite{10.1007/BFb0038202}\;
		update $L_{\textrm{uav}}^{\textrm{cand}}$ using the centor point of each minimum covering circle\;		
		update $L_{\textrm{radius}}$ using the radius of each minimum covering circle\;
		update $L_{\textrm{altitude}}$ by the relation fuction~\cite{6863654} using the corresponding radius in $L_{\textrm{radius}}$ as the input\;
		update the SINR value of each UE in $L_{\textrm{SINR}}$ by~\eqref{eq:sinr:d2u} and~\eqref{eq:sinr:g2u}\;
		\For{$i=1$ to $N$}{
			\If{$L_{\textrm{association}}[i]==-1\vee L_{\textrm{SINR}}[i]\leq\gamma_{\textrm{th}}$}{
				try to re-assocaite UE $u_i$ with another nearby UAV-BS and update $L_{\textrm{association}}$ if the SINR value exceeds $\gamma_{\textrm{th}}$\;
				\uIf{exist another one UAV can be re-assocaited by $u_i$}{
					update $L_{\textrm{association}}[i]$ and	jump to line~\ref{alg:uav_placement:20}\;
				}\Else{
					$L_{\textrm{association}}[i]\leftarrow -1$\;
				}				
			}
		}
	}
	\If{$\exists e\in L_{\textrm{association}}, e==-1$ or cannot pass any one of constraints from~\eqref{eq:system_sum_rate:c1} to~\eqref{eq:system_sum_rate:c6}}{
		$k=k+1$\;
		jump to line~\ref{alg:uav_placement:19}\;
	}
	\Return $L_{\textrm{association}}$,$L_{\textrm{uav}}^{\textrm{cand}}$, and $L_{\textrm{altitude}}$\;
	\caption{Data-Driven 3D Placement of UAV-BSs}
	\label{alg:uav_placement}
\end{algorithm2e}

\subsection{Initialization}
As the considered problem~\eqref{eq:system_sum_rate}, we can know that the system sumrate mainly depends on $N_j=\sum_{i=1}^{N}\delta_{i,j}$ and $N_G=1-\sum_{i=1}^{N}\delta_{i,j}$ which are determined by the placement of UAV-BSs. It is also similar to user association or load balancing issues of communication systems. The proposed approach uses the spatial information of UEs, UAV-BSs, and the GBS to provide an effective placement of UAV-BSs. 
Let variable $L_G=(x_G,y_G)$ record the coordinate of the GBS, a set $L_E$ store the locations of UEs, and a set $L_U$ save the locations (coordinates) of UAV-BSs.
In the system initialization stage, the system computes and store the received power of each UE from the GBS, $P_{i,G}^R=P_Ghr_{i,G}^{-\alpha}$ in a set $S_G$, where $1\leq i\leq N$. The distances from the GBS to all UEs are stored in a set $D_G$. Intuitively, UAV-BSs are used to assist the GBS, and determine the preliminary association between the GBS to each UE first. Since the number of deployed UAV-BSs and locations of deployed UAV-BSs are unknown in this stage, the interference power cannot be obtained. Instead, the initial association between the GBS to each UE will be determined by the condition $\gamma_{i,G}>\gamma_{\textrm{th}}$, where $\gamma_{i,G}$ is the SINR without considering the interference. The number of UEs having higher SINR values than $\gamma_{\textrm{th}}$ is denoted as $N_G^{\text{temp}}$. In addition, according to~\eqref{eq:rate:gbs_ue} and ~\eqref{eq:system_sum_rate:c3}, we can obtain the upper bound of $N_G$ by
\begin{equation}\label{eq:initial_n_g}
N_G^{\max}= B\text{Pr}\{\gamma_{i,j}>\gamma_{\textrm{th}}\}\log_2(1+\gamma_{\textrm{th}})/c_{\min},
\end{equation}
where $B_{i,G}=B/N_G$. Hence, we select $N_G=\min(N_G^{\text{temp}},N_G^{\max})$ UEs with the top values of the SINR to associated with the GBS.

After finishing initial association of the GBS, the system can determine the a preliminary number of UAV-BSs for clustering UEs which is obtained by
\begin{equation}\label{eq:initial_value_of_k}
k=\lceil(N-N_G)c_{\min}/\hat{C}_{\max}\rceil,
\end{equation}
where $\hat{C}^{\max}=\max\{\min\{\hat{C}_j,C_j\}|j=1,2,\dots,k\}$ is the maximum achievable data rate through the back-haul link of each UAV. Note that $\hat{C}_j$ and $C_j$ are derived by~\eqref{eq:backhaul_rate:uav} and~\eqref{eq:sum_rate_constrait:uav_to_all_ue} in a ideal case without fading and interference.

\subsection{User Association Clustering}
In the user association stage, the system makes each UE be associated with a least one UAV-BS in a best-effort manner. For UE $u_i$, $\gamma_{i,j}$ must be larger than the given threshold $\gamma_{\textrm{th}}$ so that $u_i$ can be associated with UAV-BS $U_j$. In general, the allocated data rate $c_i$ and SINR $\gamma_{i,j}$ of $u_i$ increase when the distance $r_{i,j}$ between $u_i$ and $U_j$ decreases. We consider a variation of weighted assignment problem, \emph{Capacitated Clustering Problem} (CCP)~\cite{MULVEY1984339}. The CCP is an $\mathcal{NP}$-complete decision problem and can be defined as follows.

Given a set of $N$ UEs and a set of	$k$ UAV-BS ($k<N$), let $r_{i,j}$ be the horizontal distance between UE $u_i$ and UAV-BS $U_j$ (cluster centroid), $c_{i,j}$ be the allocated data rate of UE $u_i$, $\hat{C}_j$ be the back-haul constraint of UAV-BS $U_j$, and then find $k$ disjoint subsets of UEs so that the total horizontal distance value of selected UEs is a minimum and each subset can be assigned to a different UAV-BS whose back-haul constraint is no less than the total horizontal distance value of UEs in the subset. Formally,
\begin{align}\label{eq:ccp}
\min&\enspace\sum_{j=1}^{k}\sum_{i=1}^{N}r_{i,j}\delta_{i,j},&\tag{P2}\\
s.t.&\enspace \sum_{i=1}^{N}c_{i,j}\delta_{i,j}\leq \hat{C}_j, \quad j=1,2,\dots,k,\nonumber\\
&\enspace \sum_{j=1}^{k}\delta_{i,j}=1, \qquad\quad i=1,2,\dots,N,\nonumber\\
&\enspace \sum_{j=1}^{k}\beta_j=k, \qquad\quad\: j=1,2,\dots, k,\quad j=1,2,\dots,k,\nonumber\\
&\enspace \delta_{i,j},\beta_j\in\{0,1\}, \quad\:\:\: i=1,2,\dots, n,\quad j=1,2,\dots,k,\nonumber
\end{align}
where $\beta_j$ indicates whether UAV-BS $U_j$ is deployed or not and 
\begin{equation*}\label{p:gap:eq2}
\delta_{i,j}=\begin{cases}
1, & \text{if UE $u_i$ is assigned to UAV-BS $U_j$;}\\
0, & \text{otherwise}.
\end{cases}
\end{equation*}

According to the problem~\eqref{eq:ccp}, we can know that the clustering technologies can be used to deal with the user association problem. Note that $r_{i,j}$ in~\eqref{eq:ccp} represents the cost function for the clustering. We can substitute a customized cost function for $r_i$ to obtain a different clustering result. In this stage, we adopt a \emph{balanced $k$-means clustering}~\cite{AAAI1816711} in our proposed procedure.

\subsection{Re-association and Placement Refinement}
After the association stage, the system get $k$ centroid points of the generated clusters. If we treat the horizontal coverage of each UAV-BS mapping to the ground as a ideal circle and directly deploy each UAV-BS to the centroid point of each cluster, the horizontal coverage radius of each UAV-BS will be the horizontal distance from the centroid point to the furthest UE of each cluster. However, the system using such placement will cause a large overlapping coverage area. If the overlapping coverage area becomes higher, it means that the distances between different deployed UAV-BSs become short. Such placement may leads serious co-channel interference between the UAV-BSs. To alleviate the effect of the co-channel interference, the first task of this stage, placement refinement, will be used to refine the 2D location and coverage radius of each UAV-BS. The procedure of the placement refinement solves the minimum covering circle problem~\cite{10.1007/BFb0038202} in linear time. After obtaining the minimum covering circle of each cluster, the system recognized it as the candidate coverage of each UAV-BS and the center of each minimum covering circle will be the 2D candidate location of each UAV-BS.

The system then compute and record the SINR value of each UE using the information of candidate coverage and 2D candidate location of each UAV-BS. Since the above balanced $k$-means clustering and refinement does not handle the communications constraints yet, we need to check whether the demand of each clustered UE on data rates can be satisfied in this stage. If not, it means that some UEs are too far away from its associated UAV-BS and the SINR of received signals can not exceed the threshold. In such a case, this kind of UEs may be re-associated with another nearby UAV-BS and then get the satisfied data rate. Hence, the second task of this stage is to check the communications constraints of each UE and re-associate all the unsatisfied UEs.

The last task of this stage is to judge whether the obtained candidate placement is valid by checking the existence of the unsatisfied UEs. If any unsatisfied UEs exist, it means that the obtained candidate placement is invalid and the value of $k$ may be too small to satisfied the UE demand in the considered scenario.
The system will thereby do the whole procedure of this stage repeatedly with $k=k+1$ until the obtained candidate placement is valid.

\section{Simulation Results}
\label{sec:simulation}
The simulation including all compared approaches are implemented in MATLAB R2017b. The simulation program is executed on a Windows 10 server with an Intel(R) Core(TM) i7-7700 CPU @ 3.60GHz and 8GB $\times$ 2 memory. We use 100 different artificial datasets as the input spatial information and each dataset contains $600$ to $1,300$ UE locations which are arbitrarily distributed over a $1,200\times 1,200$ m$^2$ area. We consider the urban scenario and its environmental parameters are $(a,b,\eta_{LoS},\eta_{NLoS}) = (9.61,0.16,1,20)$ given by~\cite{7744808}. 
We assume the maximum allowable path-loss of the UAV-BS to the UE link is $L_{h_j,r_{i,j}}^{\max}=119$ (dB). 
The other important simulation parameters and predefined constraints are presented in Table~\ref{simulation_parameters}. 
\begin{table}[!t]
	\renewcommand{\arraystretch}{1.3}
	\caption{Simulation Parameters}
	\label{simulation_parameters}
	\centering
	\scriptsize
	\begin{tabular}{|p{.95cm}|c||p{.95cm}|c||p{.95cm}|c|}
		\hline
		\textbf{Parameter} & \textbf{Value} & \textbf{Parameter} & \textbf{Value} & \textbf{Parameter} & \textbf{Value}\\
		\hline
		$P_G$ & $40$ dBm & $P_j$ & $20$ dBm & $P_G^{\textrm{mmWave}}$ & $30$ dBm\\
		\hline
		$\alpha$ & $6.5$ & $c_{\min}$ & $10^6$ bps & $N_{0}$ & $-174$ dBm/Hz \\
		\hline
		$h_{\min}$ & $20$ m & $f_c$ & $2$ GHz & $f_c^\textrm{mmWave}$ & $28$ GHz \\		
		\hline
		$h_{\max}$ & $400$ m & $B$ & $20$ MHz & $B^\textrm{mmWave}$ & $20*100$ MHz \\		
		\hline
		$\gamma_{\textrm{th}}$ & $5$ dB & $\gamma_{\textrm{th}}^\textrm{mmWave}$ & $30$ dB  &  &  \\		
		\hline
	\end{tabular}
	\vspace{-5pt}
\end{table}

\begin{figure}[!t]
	\centering
	\includegraphics[width=0.4\textwidth]{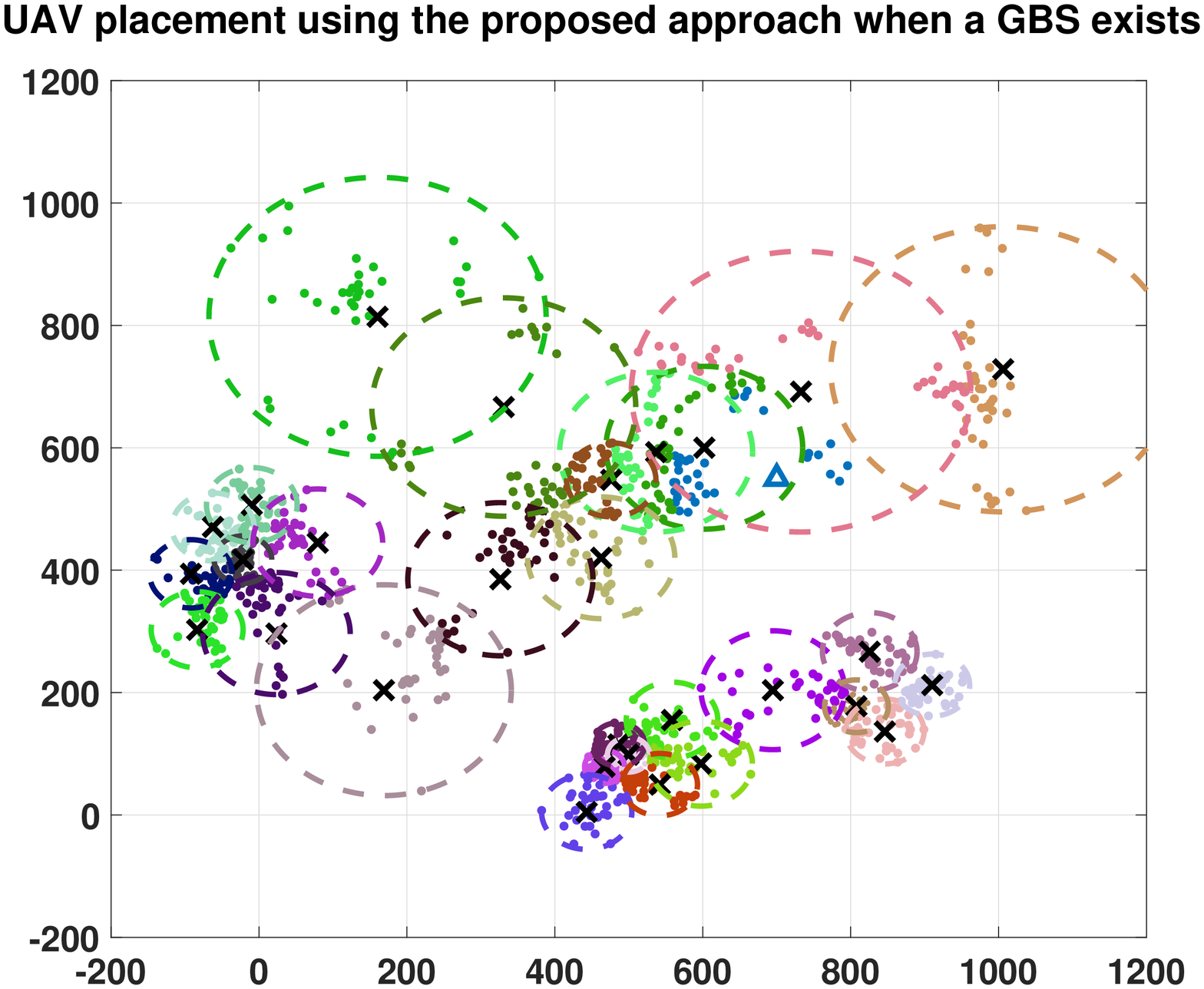}
	\caption{The placement result of the proposed approach ($k=39$) when $N=1100$.}
	\label{fig:placement_result}
	\vspace{-5pt}
\end{figure}

\begin{figure}[!t]
	\vspace{-5pt}
	\centering
	\includegraphics[width=0.47\textwidth]{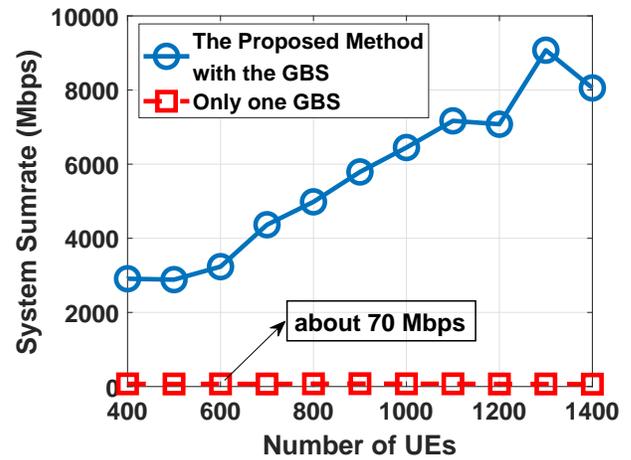}
	\caption{The improvement of system sumrate.}
	\label{fig:sumrate_comparison} 
	\vspace{-10pt}
\end{figure}

The simulation results are depicted in Fig.~\ref{fig:placement_result} and Fig.~\ref{fig:sumrate_comparison}. We choose one of input location data set to show the placement results of the proposed approach in Fig.~\ref{fig:placement_result}. Note that the blue triangle is the GBS, black crosses are UAV-BSs, small dots are UEs, and each dashed-circles is the coverage of the corresponding UAV-BS. In this scenario, some very high dense flash crowd events occurs around the following coordinates: $(0,400)$, $(500,500)$, $(500,100)$, and $(820,200)$. We can see that the proposed approach can provide a placement with small overlapping coverage area. Unlike the conventional $k$-means++ which cannot determine the value of $k$ by the algorithm itself, the proposed approach will automatically determine an appropriate initial value of $k$ according to the input spatial information, environmental parameters, and communications constraints. In this way, the system can save a lot of computational costs (time and energy) on checking the impossible value of $k$. 

The result in Fig.~\ref{fig:sumrate_comparison} indicates that the proposed approach can provide a significantly improved system sumrate comparing with the one without using any UAV-BSs while serving different numbers of UEs. It also shows that the considered systems has a bottleneck of sumrate when more than 1300 UEs locate in the serving area $1,200\times 1,200$ m$^2$. If $N>1300$, the system needs more UAV-BSs to fulfill the minimum data rate requirement of each UE. However, such a situation leads to significant co-channel interference and thus reduce the system sumrate.

\section{Conclusion}
\label{sec:conclusion}
In this paper, we discuss how to deploy multiple UAV-BSs for serving arbitrary flash crowds. The proposed data-driven 3D placement algorithm can automatically determine the appropriate number, location, altitude, and coverage of each UAV-BS and then place the UAV-BSs in polynomial time. According to the simulation results, the proposed approach improves the system sumrate as well as guarantees the minimum data rate requirement of each UE effectively. 

\bibliographystyle{IEEEtran}
\bibliography{IEEEabrv,reference}

\end{document}